\documentclass[aip,apl,twocolumn,showpacs,reprint]{revtex4-1}
\usepackage{graphicx}
\usepackage{dcolumn}
\usepackage{bm}
\usepackage{amssymb}
\DeclareGraphicsExtensions{.pdf,.eps,.png,.jpg,.mps,.tif}

\newcommand{\beq}{\begin{equation}}  
\newcommand{\eeq}{\end{equation}}  
\newcommand{\beqa}{\begin{eqnarray}}  
\newcommand{\eeqa}{\end{eqnarray}}

\begin{document}

\title{Probing a Single Nuclear Spin in a Silicon Single Electron Transistor}

\author{F. Delgado$^{(1)}$, R. Aguado$^{(2)}$, and J. Fern\'andez-Rossier$^{(1,3)}$}
\affiliation{$^{(1)}$ International Iberian Nanotechnology Laboratory (INL),
Av. Mestre Jos\'e Veiga, 4715-330 Braga, Portugal
\\ $^{(2)}$ Instituto de Ciencia de Materiales de Madrid
(ICMM-CSIC), Cantoblanco, 28049 Madrid, Spain
\\ $^{(3)}$ Departamento de F\'isica Aplicada, Universidad de Alicante, 03690 San Vicente del Raspeig, Spain}

\date{\today}

\begin{abstract}

We study single electron transport across a single Bi dopant in a Silicon Nanotransistor to assess  how the strong hyperfine coupling  with the  Bi nuclear spin $I=9/2$ affects  the transport characteristics of the device.   In the  sequential tunneling regime we find that at,  temperatures in the range of $100 mK$, $dI/dV$ curves reflect the zero field hyperfine splitting as well as  its evolution under an applied magnetic field.  Our non-equilibrium quantum simulations show that nuclear spins can be partially polarized parallel or antiparallel to the electronic spin just tuning the applied bias.

\end{abstract}

 \maketitle


The amazing progress both in the silicon processing technologies and in the miniaturization of silicon based transistors
 has reached the point where single-dopant transistors have been demonstrated.\cite{Sellier_Lansbergen_PRL_2006,Pierre_Wacquez_naturenano_2009,
Lansbergen_Tettamanzi_NanoLetters_2009, Tan_Chan_NanoLetters_2010,
Golovach_Jehl_PRB_2011,Fuechsle_Miwa_natnanotech_2012,Tettamanzi_Verduijn_prl_2012}
 Whereas this progress has been fueled by the development of classical computing architectures,  it might also be used for quantum computing. In this regard, the electronic and nuclear spins of single donors in silicon  are very promising building blocks for  
 quantum computing.\cite{Kane_Nature_1998,DiVincenzo_Bacon_nature_2000,Ladd_Jelezko_nature_2010}
 Progress along this direction makes it necessary to implement single spin readout schemes both for electronic and nuclear spins. 
 Single electronic spin readout has been demonstrated, both in GaAs quantum dots as well as in P doped Silicon Nanotransistors.\cite{Elzerman_Hanson_nature_2004,Morello_Pla_nature_2010}

The readout of the quantum state of a single nuclear spin,
  much more challenging, has  been demonstrated for NV centers in diamond taking advantage of single spin optically detected magnetic resonance afforded by the extraordinary properties of that system.\cite{Neumann_Beck_science_2010}  Single nuclear spin readout
   with either optical\cite{Fu_Ladd_prb_2004}
   or a combined electro-optical techniques\cite{Sleiter_Kim_newjphys_2010}  has been proposed, but remains to be implemented.
Here we explore the  electrical readout  of a  single nuclear spin, more suitable for   an indirect band-gap host  like Si. A preliminary step   is to construct a circuit whose transport is affected by the quantum state of the nuclear spin.  There is ample experimental evidence of the mutual influence of many nuclear spins  and transport electrons in  III-V semiconductor quantum dots in the single electron transport regime.\cite{Petta_Taylor_prl_2008,Reilly_Taylor_science_2008,Foletti_Bluhm_natphys_2009,Kobayashi_Hitachi_prl_2011}  In particular,   Kobayashi {\em et al.}  have reported hysteresis in the $dI/dV$ upon application of magnetic fields, reflecting the realization of different ensemble of nuclear  states coupled to the electronic spin via hyperfine coupling.\cite{Kobayashi_Hitachi_prl_2011}
  
\begin{figure}
\includegraphics[height=0.4\linewidth,width=0.85\linewidth,angle=0]{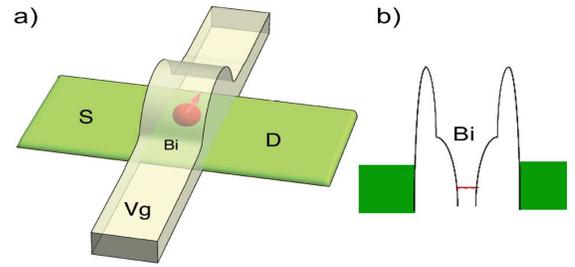}
\caption{ \label{fig0} (color online) a) Scheme of the Si:Bi FinFET nanotransistor.
b) Trapping Coulomb potential of the Bi dopant an single energy level participating in the transport.  
}
\end{figure}

Here we propose a device where a single nuclear spin is probed in single electron transport.  We model the single electron transport in a silicon nanotransistor such that, in the active region,  transport takes place through  a single Bi dopant, see Fig.~\ref{fig0}.  We show that, at sufficiently low temperatures, the $dI/dV$ curves of this device
probe the hyperfine structure of the dopant level. In turn,  the occupations of the nuclear spin states are affected by the transport electrons.
Whereas single dopant transistors have  been demonstrated for single P, As and B, in Si,
\cite{Lansbergen_Tettamanzi_NanoLetters_2009,Tan_Chan_NanoLetters_2010,Morello_Pla_nature_2010,Fuechsle_Miwa_natnanotech_2012} we choose Bi because it has a much larger hyperfine splitting,\cite{George_Witzel_prl_2010,Morley_Warner_natmat_2010,Mohammady_Morley_prl_2010} due to both a larger nuclear spin $I=9/2$ and a larger hyperfine coupling constant ($A\approx 6.1\mu$eV). 
The zero-field splitting of the Bi donor level is given by $5A$ and has been observed by electron spin resonance\cite{George_Witzel_prl_2010,Morley_Warner_natmat_2010,Mohammady_Morley_prl_2010} and in photoluminescence experiments with many dopants.\cite{Sekiguchi_Steger_prl_2010}

%
%

We consider the sequential transport regime, where the occupation of the donor level fluctuates between $q=0$ and $q=1$. 
In the $q=0$ state, the nuclear spin interacts only with the external field.  In the $q=1$ state, the electron and the nuclear spin are hyperfine coupled. The Hamiltonian that describes both states reads \cite{George_Witzel_prl_2010,Morley_Warner_natmat_2010,Mohammady_Morley_prl_2010,Delgado_Rossier_prl_2011}
\beqa
{\cal H}=q\left(\epsilon_d+ eV_G +A\vec{S}\cdot \vec{I}+\hbar \omega_e S_z\right)+\hbar \omega_N I_z, 
\eeqa
 where $\epsilon_d$  is the donor energy level with respect to the Fermi energy, which we take as $E_F=0$, and
$V_G$ denotes an external gate voltage.    
We assume that valley degeneracies of the donor level are split-off and neglect the valley degree of freedom. 
The third term
 is the hyperfine coupling, and the last two, where $\hbar \omega_e=g_e\mu_B B_z$ and $\hbar \omega_N=g_n\mu_N B_z $, correspond to the electron and nuclear Zeeman terms, with $g_e$ ($g_n$) the electron (nuclear) g-factors and $\mu_B$ ($\mu_N$) the Bohr (nuclear) magneton.
In equilibrium, i.e., at zero bias, the occupation of the dopant level depends on the value of the addition energy, which ignoring the Zeeman terms and the tiny correction due to the hyperfine coupling, is given by $\varepsilon_0(V_G)\equiv \epsilon_d+ eV_G$.

\begin{figure}
\includegraphics[width=0.9\linewidth,angle=0]{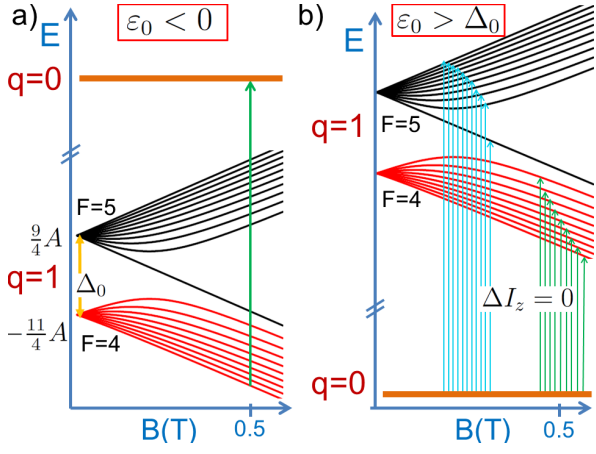}
\caption{ \label{fig2} (color online) Scheme of the  current-induced allowed transition for the a) $q=1$ charged system  and b) $q=0$ uncharged system. It has been assumed that $\hbar \omega_N \ll k_B T\ll \hbar\omega_e \lesssim \Delta_0$.
}
\end{figure}

We denote the $q=0$ eigenstates as $|m\rangle$. Their energies read as $\epsilon_m\equiv  \hbar \omega_N I_z$.
The eigenenergies and eigenvectors of $q=1$  are denoted by $\epsilon_M$ and $|M\rangle $.
The $q=1$
  zero-field Hamiltonian $ A\vec{I}\cdot\vec{S}$ can be diagonalized in terms of the total angular operator $F$, resulting in two multiplets (F=4, F=5)  with energies $E_{F=4}=-11A/4$ and  $E_{F=5}=9A/4$, and a zero-field  splitting $\Delta_0=5A\approx 30\mu$eV.
   At finite magnetic field, the exact eigenvalues of ${\cal H}$ can also be calculated  analytically.\cite{Mohammady_Morley_prl_2010}  The corresponding energy levels are shown in Fig.~\ref{fig2}.

The tunneling Hamiltonian  between the single Bi dopant level and the source and drain  electrodes 
reads as
\begin{equation}
\label{htun}
{\cal H}_{tun}= \sum_{\lambda\sigma} V_\lambda \left(d_\sigma^\dag c_{\lambda\sigma} +h.c\right),
\end{equation}
 where operator  $c_{\lambda,\sigma}$ annihilates an electron with spin $\sigma$ and orbital quantum number $\lambda\equiv \eta,\vec{k}$, with wave vector $\vec{k}$ and electrode index $\eta=S,D$, while operator  $d_\sigma$ annihilates a spin $\sigma$ electron in the dopant level. 
 The scattering rate for the tunneling process, ignoring the hyperfine coupling, is given by $\Gamma_0^{\eta}=\frac{2\pi}{\hbar}|V_{\eta}|^2\rho_\eta$, where $\rho_{\eta}$ is the density of states of the electrode.   
Our model  is  very similar to the one used to describe single electron transport through a quantum dot exchanged coupled to a single Mn atom.\cite{Efros_Rashba_prl_2001,Rossier_Aguado_prl_2007}

The dissipative dynamics of the electro-nuclear spin system 
 under the influence of the coupling to the electrodes is described  by a Bloch-Redfield (BR)  master equation.\cite{Cohen_Grynberg_book_1998,Rossier_Aguado_prl_2007}
The coupling to the reservoir, given by the tunneling Hamiltonian,  involves transitions between the $q=0$ and $q=1$ manifolds.  The corresponding
transition rates  are be calculated using the Fermi golden rule with ${\cal H}_{tun}$ as the perturbation:\cite{Rossier_Aguado_prl_2007}
\beqa
\label{ws}
\Gamma^{\eta}_{m,M}&=&\Gamma_0^\eta\sum_\sigma \left| \langle M|I_z(m),\sigma\rangle\right|^2,
\eeqa
where $|I_z,\sigma\rangle \equiv |I_z\rangle \otimes|\sigma\rangle$.
 In the following we take the applied bias convention $\mu_S-\mu_D=eV$, with $\mu_S=eV/2$ and $\mu_D=-eV/2$. For a given temperature, bias and gate voltages and Hamiltonian parameters, we 
obtain the steady state solution  of the master equation, ignoring
 the effect of the fast-decaying coherences. This yields the steady state occupations $P_m(V)$ and $P_M(V)$.
  
We consider the sequential tunneling regime, in which the energy level broadening induced by coupling to the electrodes $\Gamma_0$ is small, $\hbar \Gamma_0\ll k_B T$. 
This also justifies the  markovian approximation implicit in the Bloch-Redfield master equation.  In this regime, current flows when the bias enables charge fluctuations of the dopant level.  The steady state current corresponding to electrons flowing from the source electrode  to the dopant level
is given by
\beqa
I&=&e\sum_{m,M}\Big\{P_{m}(V)  f_S(\Delta_{M,m})\Gamma^{S}_{m,M}
\crcr
 &&-P_M(V) \left[1-f_S(\Delta_{M,m})\right]\Gamma^{S}_{m,M}
 \Big\},
\label{ist}
\eeqa
where $\Delta_{M,m}=\epsilon_{M}-\epsilon_{m}$ and 
$f_S(\epsilon)=f(\epsilon-\mu_S)$ is the Fermi function relative to the chemical potential of the $S$ electrode.  
The first term in the right hand-side of Eq.($4$) represents the electrons flowing from  the $S$ electrode to the empty Bi, while the second one corresponds to electrons flowing from the $q=1$ Bi to the $S$ electrode. In steady state, the continuity equation ensures that current  between the dopant and the drain is the same than the source-dopant current. 

\begin{figure}[t]
\includegraphics[width=1.\linewidth,angle=0]{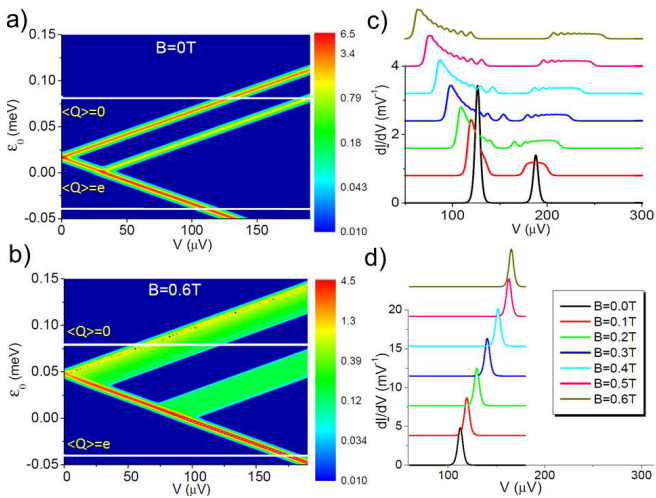}
\caption{ \label{fig1} (color online) a) and b) Contour plot of the $d\underline{I}/dV$ vs. applied bias $V$ and on-site energy $\varepsilon_0$ at zero magnetic field (left) and $B=0.6$T (right). c) and d) 
Conduction spectrum  $d\underline{I}/dV$  as a function of applied bias for different magnetic fields at $\varepsilon_0=-0.4\mu$eV and $\varepsilon_0=0.8$meV respectively. White horizontal lines in panel a) and b) marks the values of $\varepsilon_0$ in the 2D plots c) and d). In all cases, 
$T=10$mK and $\hbar \Gamma_0=0.1\mu$eV.}
\end{figure}

Figure ~\ref{fig1}a) shows the differential conductance $d\underline{I}/dV$  map  for zero-applied magnetic field, with $\underline{I}=I/(e\Gamma_0)$ and $\Gamma_0^\eta=\Gamma_0/2$.  
At zero bias, the conductance is zero except at the special value of $V_G$ for which the addition energy vanishes.   Far from this point, the zero-bias charge of the dopant state, hereafter denoted with $q_0$,  is either $q_0=0$ or $q_0=1$.  The finite bias conductance has a peak whenever the bias  energy, $eV/2$, matches the energy difference between two states with different charge, $m$ for $q=0$ and $M$ for $q=1$,  that  are permitted by the spin selection rule implicit in  Eq.(3). 
 The height of the peak is proportional to both the non-equilibrium occupations $P_m$ and $P_M$ 
 and to the quantum mechanical matrix element $\Gamma^{\eta}_{m,M}$.  This determines the very different spectra when the zero bias charge in the dopant is $q=0$ or $q=1$. 
The width of the $dI/dV$ peaks is proportional to $k_BT$, so that 
 the $dI/dV$ spectra can resolve the hyperfine structure provided that $k_BT$ is smaller than the splitting of the levels.   
The energy differences inside the $F=4$ and $F=5$ manifolds, see Fig.~\ref{fig2}), are roughly proportional to $A$. 
Thus, while the zero-field splitting can be resolved at $T=0.3$ K, temperature must be significantly below 50 mK to resolve the finite field structure, see Fig.~\ref{fig1}c).

\begin{figure}
\includegraphics[width=0.99\linewidth,angle=0]{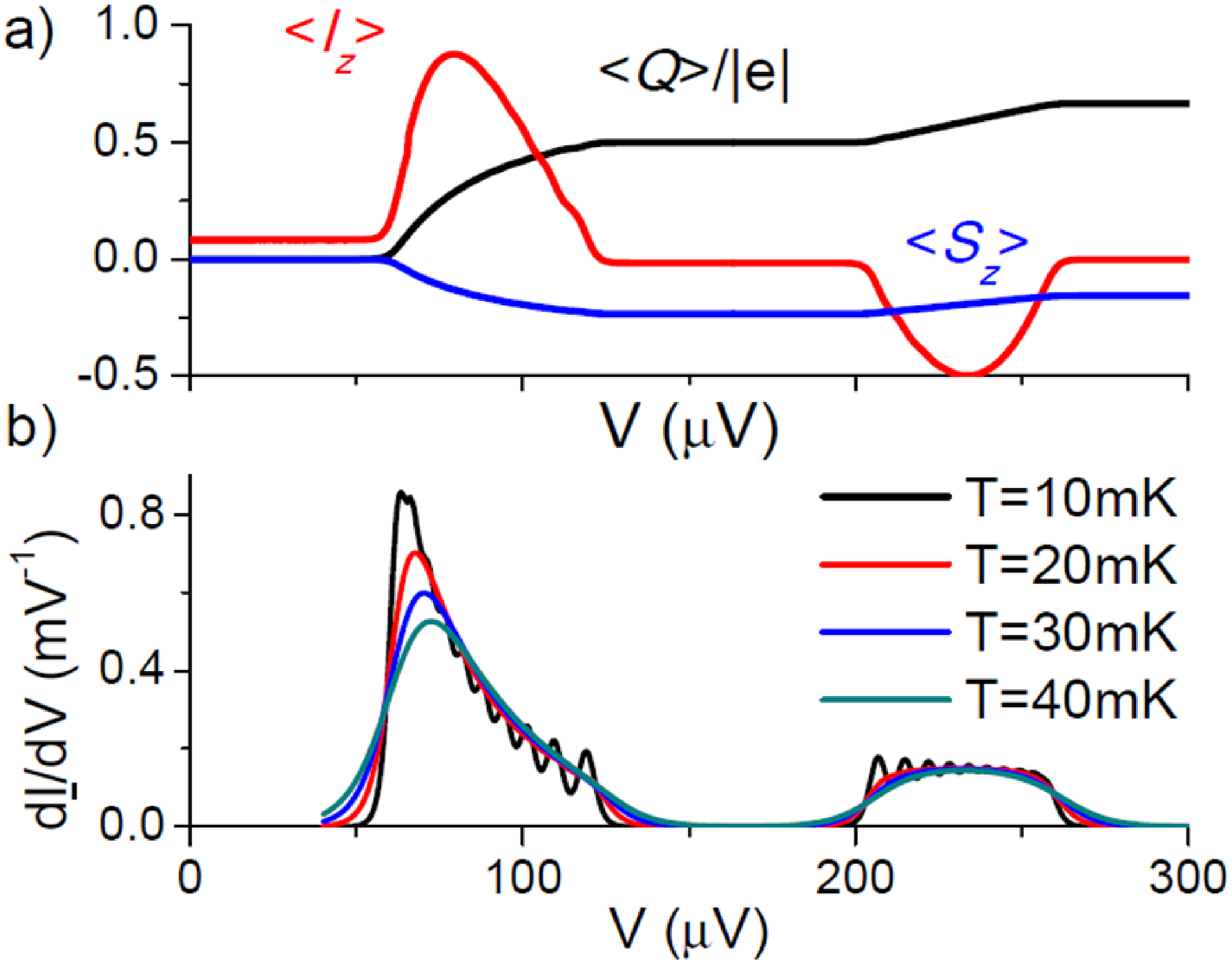}
\caption{ \label{fig3} (color online)  a) Average electronic occupation  of the Bi, $\langle Q\rangle/|e|$ (black line) and  nuclear and electronic spins, $\langle I_z\rangle$ (red line) and $\langle S_z\rangle$ (blue line), respectively. b) $d\underline{I}/dV$ vs. bias for different temperatures.
Same parameters as ~\ref{fig1}c) with $B=0.6$T. 
}
\end{figure}

 Let us consider first  the $q_0=1$ case (left panel in Fig.~\ref {fig1}). At $10$ mK only the ground state(s) is (are) occupied. Thus, a single transition is seen, from the $q=1$ to the $q=0$ states. As the magnetic field is ramped, the energy of the transition increases, reflecting the electronic Zeeman shift.  In contrast,  in the $q_0=0$ case (right panel in Fig.~\ref{fig1}), all the   Zeeman split nuclear levels  are equally populated, even down to  mK temperatures.  Spin conservation selection rule implicit in Eq.~(2) 
connects these 10 quasi-degenerate states of the $q=0$  manifold  to the hyperfine spin-split levels of the $q=1$ manifold with different energies. As a result, the $dI/dV$ curve reveals 2 peaks at zero field, reflecting the splitting between the $F=4$ and $F=5$ states. At higher fields,  the two zero-field peaks split in up to  10  peaks, that  can be resolved at low enough  temperature [see Figs.~\ref{fig1}c) and ~\ref{fig3}b)].
 
 Interestingly,  the application of a bias to the $q_0=0$ state, for which the nuclear spin states are randomized, can result
 in a finite average nuclear magnetic moment.  We show this in  Fig.~\ref{fig3}a) for finite $B$.  At zero bias, the charge of the dopant level is $q_0=0$, and the nuclear spins are randomized. When the bias hits the addition energy a selective depopulation of a given $I_z$ level of the $q=0$ manifold starts, in favor of a $q=1$ state that mixes  the $I_z$ and $I_z \pm 1$ components, resulting in a net accumulation of nuclear spin.   When all the transitions  to the $F=4$ manifold are allowed, the nuclear spin vanishes again. Then,  when the bias permits the transitions to the $F=5$ manifold, the nuclear spin accumulation starts in the opposite direction.
 Thus, when $|eV/2|$ matches the center of the $F=4$ multiplet, see Fig.~\ref{fig3}a), the nuclear spins tend to align antiparallel to the electronic spin. Then, when $|eV/2|$ reaches the center of the $F=5$ multiplet, the nuclear spins prefer aligning parallel to the electronic spin. 

Whereas all our results discussed so far refer to steady state conditions, it is worth pointing out that there are two very different time scales in the dynamics of the system.
Whereas the charge equilibrates in the dopant level in a time scale set by $1/\Gamma_0$, the nuclear spin relaxation, dominated by many events of hyperfine exchange with the electronic spin and subsequent recharging of the Bi, \cite{Besombes_Leger_prb_2008}  takes place at a much longer time scale, hundreds of time larger than $1/\Gamma_0$, but still much shorter than the intrinsic $T_1$ time of the nuclear spin.  Thus, charge fluctuations in the Bi induce nuclear spin relaxation.\cite{Besombes_Leger_prb_2008}

 We finally discuss the experimental feasibility of our proposal with state of the art techniques. First, according to our simulations, see Fig.~\ref{fig3}b), the finite field hyperfine splitting is resolved at 10 mK but not a 20 mK.  At 40 mK the 2 humps associated to the $F=4$ and $F=5$ manifolds are clearly resolved.  
Keeping the transport in the sequential tunneling regime requires that $\hbar \Gamma_0 \ll k_B T$, which at 10mK,  translates into $I\ll 200$pA. This is within reach of experimental setups.\cite{Kobayashi_Hitachi_prl_2011,Petta_Taylor_prl_2008,Baugh_Kitamura_prl_2007,Morello_Pla_nature_2010,Jespersen_Rasmussen_natphys_2011} 

  In conclusion, we have studied the single electron transport spectroscopy of the hyperfine structure of a  Bi dopant in a silicon nanotransistor.  We have shown that, at sufficiently low temperatures, and when the dopant is ionized with a gate,  the $dI/dV$ corresponding to sequential transport can resolve the hyperfine spectrum of the electron in the donor level.   In addition, the non-equilibrium transport at finite field results in a hyper polarization of the nuclear spin state, or nuclear spin accumulation.    These results are different from our previous work, where we considered the same system in a different transport regime, cotunneling, and we showed that inelastic cotunneling of the dopant in the $q=1$ state could also resolve the hyperfine spectrum and drive the nuclear spin states out of equilibrium.\cite{Delgado_Rossier_prl_2011}  Future work should determine how, in the cotunneling regime, the appearance of the Kondo effect \cite{Tettamanzi_Verduijn_prl_2012} competes with the reported effect.

    This work has been financially supported by MEC-Spain (Grant Nos. FIS2010-21883-C02-01, FIS2009-08744,  and CONSOLIDER CSD2007-0010) as well as Generalitat Valenciana, grant Prometeo 2012-11.

%
%






\end{document}